\newcommand{\dev}[1]{}
\newcommand{\rel}[1]{#1}

\documentclass[twocolumn,prl,showpacs]{revtex4}

\usepackage[latin1]{inputenc}
\usepackage{times}
\usepackage{amsfonts}
\usepackage{amsmath}
\usepackage{amssymb}
\usepackage{color}
\usepackage{graphics}
\usepackage{epsfig}
\usepackage{psfig}
\usepackage{bm}

\newcommand{\bc}{\begin{center}}
\newcommand{\ec}{\end{center}}

\newcommand{\beq}{\begin{equation}}
\newcommand{\eeq}{\end{equation}}
\newcommand{\beqar}{\begin{eqnarray}}
\newcommand{\eeqar}{\end{eqnarray}}
\newcommand{\beqars}{\begin{eqnarray*}}
\newcommand{\eeqars}{\end{eqnarray*}}

\newcommand{\Figref}[1]{Fig.~\ref{#1}}

\newcommand{\Eqref}[1]{Eq.~(\ref{#1})}

\newcommand{\Eref}[1]{(\ref{#1})}

\newcommand{\topage}[1]{}

\newcommand{\matrixer}[1]{#1}
\newcommand{\vectorer}[1]{\bm{#1}}
\def\eps{\varepsilon}
\def\Deps{\Delta\varepsilon}
\def\deps{\delta\varepsilon}
\def\Dphi{\Delta\phi}
\def\dphi{\delta\phi}
\def\Dvphi{\Delta\varphi}
\def\DPhi{\Delta\Phi}

\def\iU{U^{-1}}

\begin{document}

\title{Breaking Synchrony by Heterogeneity in Complex Networks}

\author{Michael \surname{Denker}}
\author{Marc \surname{Timme}}
\author{Markus \surname{Diesmann}}
\author{Fred \surname{Wolf}}
\author{Theo \surname{Geisel}}
\affiliation{Max-Planck-Institut f\"ur Str\"omungsforschung and Fakult\"at f\"ur Physik, Universit\"at G\"ottingen, 37073 G\"ottingen, Germany}
%\date{1.1.1111}

\begin{abstract}
For networks of pulse-coupled oscillators with complex connectivity, we demonstrate that in the presence of coupling heterogeneity precisely timed periodic firing patterns replace the state of global synchrony that exists in homogenous networks only. With increasing disorder, these patterns persist until they reach a critical temporal extent that is of the order of the interaction delay. For stronger disorder these patterns cease to exist and only asynchronous, aperiodic states are observed. We derive self-consistency equations to predict the precise temporal structure of a pattern from the network heterogeneity. Moreover, we show how to design heterogenous coupling architectures to create an arbitrary prescribed pattern.
\end{abstract}

\maketitle

Understanding how the structure of a complex network \cite{cnpaper} determines its dynamics is currently in the focus of research in physics, biology and technology \cite{dynnet}. Pulse-coupled oscillators provide a paradigmatic class of models to describe a variety of networks that occur in nature, such as populations of fireflies, pacemakers cells of the heart, earthquakes, or neural networks \cite{pcoex,mirollos}. Synchronization is one of the most prevalent kinds of collective dynamics in such networks \cite{synchimportant}. Recent theoretical studies, which analyze conditions for the existence and stability of synchronous states, have focused on \textit{homogenous} networks with simple topologies, e.g.\ global couplings \cite{fullycon} or regular lattices \cite{lattice}. In nature, however, intricately structured and \textit{heterogenous} interactions are ubiquitous. Previously, aspects of heterogeneity have been studied mostly in globally coupled networks \cite{tsodyks,golomb,neltner}. However, for networks with structured connectivity only a few studies exist \cite{golombH,kopell} and it is still an open question, how heterogeneity influences the dynamics, in particular synchronization, in such networks.

In this Letter we present an exact analysis of the dynamics of complex networks of pulse-coupled oscillators in the presence of coupling heterogeneity. We demonstrate that the synchronous state, that exists in homogenous networks, is replaced by precisely timed periodic firing patterns. The temporal extent of these patterns grows with the degree of disorder. Patterns persist below a critical strength of the disorder beyond which only asynchronous, aperiodic states are observed. We show how this transition is controlled by the interaction delay. Simple criteria for the stability of firing patterns are derived. Furthermore, an approach is presented to predict the relative timings of firing events from self-consistency conditions for the phases in a given network. Conversely, any prescribed pattern can be created by designing a heterogenous coupling architecture in a network of specified connectivity.

\begin{figure}
\centering
\dev{\epsfig{file=figs/fig1ext.eps, width=0.9\columnwidth}}
\rel{\epsfig{file=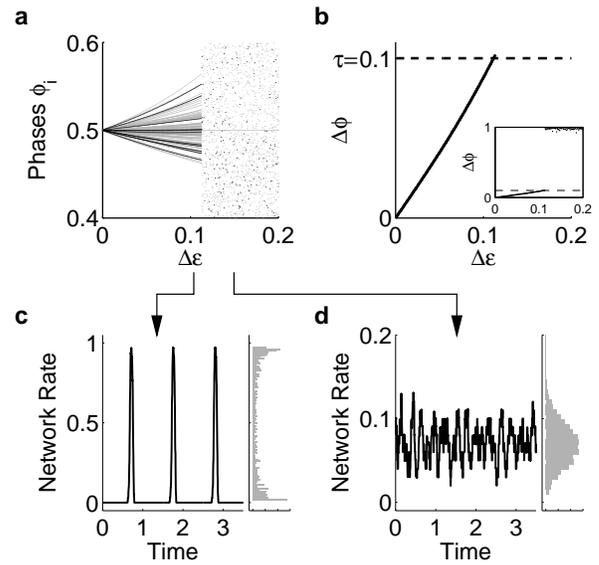, width=0.9\columnwidth}}
\caption{Heterogeneities in a random network ($N=100$, $p=0.4$, $b=2$, $\hat\eps=-0.1$, $\tau=0.1$) induce desynchronization followed by a transition to an asynchronous, aperiodic state. (a) Relative phases (ten selected phases shown in black) for different disorder strengths $\Deps$ (fixed $J$). (b) Increase of the extent $\Dphi$ of the pattern shown in (a), until the pattern disappears at $\Dphi_c\approx\tau$ (dashed line at $\Dphi=\tau$). Inset: Onset of asynchronous state ($\Dphi\approx 1$) for large disorder. (c),(d) Instantaneous network rate of (c) a periodic ($\Deps=0.10$) and (d) an aperiodic ($\Deps=0.15$) state in real time computed using a triangular sliding time window. Grey histograms on the right of (c) and (d) indicate rate distribution (excluding peak at zero network rate in (c)).}
\label{fig:1}
\end{figure}

Consider a system of $N$ pulse-coupled oscillators that interact via directed connections. A matrix $\matrixer{C}$ defines the connectivity of the network, where $C_{ij}=1$ if a connection from oscillator $j$ to $i$ exists, and $C_{ij}=0$ otherwise. The number of inputs $k_i:=\sum_jC_{ij}$ to every oscillator $i$ is non-zero, $k_i\ge 1$, and no further restriction on the network topology is imposed. In simulations we consider random graphs in which every directed connection is present with probability $p$.

The state of an individual oscillator $j$ is represented by a phase variable $\phi_j$ that increases uniformly in time \cite{mirollos},
\beq
\mathrm {d}\phi_j /\mathrm {d}t=1.
\label{eq:uniformphaseadv}
\eeq
Upon crossing a firing threshold, $\phi_j(t_f)\ge 1$, at time $t_f$ an oscillator is instantaneously reset to zero, $\phi_j(t_f^+)=0$, and a pulse is sent. After a delay time $\tau$, this pulse is received by all oscillators $i$ connected to $j$ (for which $C_{ij}=1$) and induces an instantaneous phase jump
\beq
\phi_i((t_f+\tau)^+)=\iU(U(\phi_i(t_f+\tau)+\eps_{ij})).
\eeq
Here, $\eps_{ij}$ are the coupling strengths from $j$ to $i$, which are taken to be either purely inhibitory (all $\eps_{ij}<0$) or purely excitatory (all $\eps_{ij}>0$). The interaction function $U$ is monotonically increasing, $U'>0$, and represents the subthreshold dynamics of individual oscillators. We consider functions with a curvature of constant sign, i.e.\ $U''(\phi)>0$ or $U''(\phi)<0$ for all $\phi$. This choice characterizes a  wide range of biologically relevant behaviors, such as neural membrane potential dynamics. The qualitative behavior of systems with a more complicated interaction function may be derived from the results presented here (cf.~\cite{denker}). In simulations, we use $U(\phi)=\ln(1+(e^b-1)\phi)/b$, where $b$ parameterizes the curvature of $U$. 

We briefly consider homogenous networks with individual coupling strengths $\eps_{ij}=\hat\eps_{ij}$ that are normalized
\beq
\sum_{j=1}^N\hat\eps_{ij}=\hat\eps
\label{eq:eps_norm}
\eeq
such that every oscillator $i$ receives the same total input. Such networks exhibit a synchronous state, defined by
\beq
\phi_i(t)=\phi(t)
\label{eq:synstate}
\eeq
for all $i$. The stability of this state has been determined previously \cite{timme} and depends on the interplay between the sign of the coupling (excitatory or inhibitory) and the local sign of the curvature of the interaction function $U$ \cite{denker}.

\begin{figure}
\centering
\dev{\epsfig{file=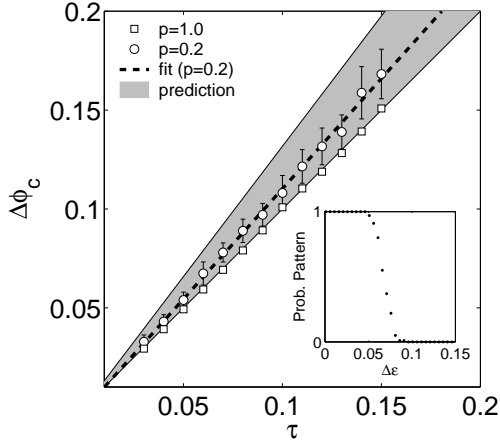, width=0.77\columnwidth}}
\rel{\epsfig{file=fig2.eps, width=0.77\columnwidth}}
\caption{The critical extent of a pattern $\Delta\phi_c$ as a function of the interaction delay $\tau$ ($N=100$, $b=2$, $\hat\eps=-0.1$). Each data point is obtained for a different heterogeneity $\matrixer{J}$ in fully connected networks ($\square$) and random networks ($\bigcirc$) of different connectivity $\matrixer{C}$ with $p=0.2$. For the random network, data are averaged over 20 trials (dashed line: fit). Grey area indicates the theoretical second-order prediction of the transition zone between $\Dphi_c^{\mathrm{glob.}}$ and $\Dphi_c^{\mathrm{max}}$. Inset: Probability that a periodic pattern emerges from synchronous initial conditions for given disorder $\Deps$ (histogram of 100 random heterogeneities and connectivities, $p=0.2, \tau=0.1$).}
\label{fig:2}
\end{figure}

Heterogeneity of the couplings is introduced through
\beq
\eps_{ij}=\hat\eps_{ij}+\Deps J_{ij}C_{ij}/k_i
\label{eq:eps_jittered}
\eeq
such that in general the total input is different for every oscillator. The matrix $\matrixer{J}$ specifies the structure of the heterogeneity where each element $J_{ij}$ is drawn randomly from the uniform distribution on $[-1,1]$. The degree of heterogeneity is quantified by the disorder strength $\Deps$. In numerical simulations, we take the homogenous part of the coupling strength to be $\hat\eps_{ij}=\hat\eps C_{ij}/k_i$.

The synchronous state \Eref{eq:synstate} of homogenous networks ($\Deps=0$) is replaced by periodic firing patterns in the presence of heterogeneity ($\Deps>0$), cf.~\Figref{fig:1}. After a discontinuous transition at a certain critical disorder $\Deps_c$ these patterns disappear and aperiodic, asynchronous states are observed \cite{howmeasure}.

The firing patterns emerging for $\Deps<\Deps_c$ are confined to a subinterval of the phase axis. The extent of a pattern
\beq
\Dphi=\max_i \phi_i - \min_i \phi_i
\label{eq:dphidef}
\eeq
(taking into account wrap-around effects at the threshold) increases as the disorder strength $\Deps$ is increased, whereas the firing order is determined by the structure $\matrixer{J}$ of the heterogeneity. If $\Deps\geq\Deps_c$ the periodic pattern is replaced either directly by an asynchronous state (\Figref{fig:1}), or it reaches this state through a sequence of states consisting of several separated clusters of firing patterns (not shown).

To understand the origin of this transition, we note that the critical disorder $\Deps_c$ is directly related to the critical pattern extent $\Dphi_c$ before breakdown (cf.\ \Figref{fig:1}(b)). A periodic pattern is always observed for $\Dphi\leq\tau$. For globally connected networks, the transition to an asynchronous state occurs at 
\beq
\Dphi_c^{\mathrm{glob.}}=\tau
\label{eq:globbound}
\eeq
(see \Figref{fig:2}). If the system is initialized with a pattern of extent $\Dphi>\Dphi_c^{\mathrm{glob.}}$, some oscillators, which we term critical, send pulses that affect oscillators that have not fired within the period considered, causing divergence from the periodic state. In networks that are not globally connected, the firing patterns may persist up to a critical extent $\Dphi_c>\Dphi_c^{\mathrm{glob.}}$ that depends on the specific network structure. For random networks, we estimate an upper bound $\Dphi_c^{\mathrm{max}}$ for $\Dphi_c$ by assuming that the firing times (i.e.~the phases) within the pattern are uniformly and independently distributed in an interval of length $\Dphi$ \cite{explainphasedist}. We first calculate the probability
\beq
P'=1-\frac{\tau}{\Dphi}-\frac 1{k+1}\Big (1-\Big (\frac{\tau}{\Dphi}\Big)^{k+1} \Big)
\label{eq:predstep}
\eeq
that a given oscillator $i$ is critical, i.e.\ that of the approximately $k=pN$ oscillators $j$ receiving input from $i$, at least one satisfies $\phi_i-\phi_j>\tau$. In a second step, we calculate the probability $P=1-(1-P')^N$
that at least one of the $N$ oscillators is critical. Setting $P=1$ guarantees the existence of a critical oscillator and results in an implicit formula for  $\Dphi_c^{\mathrm{max}}$ in terms of $N, p$ and $\tau$. To second order in $\Dphi-\tau$ we obtain
\beq
\Delta\phi_c^{\mathrm{max}}=(1+\sqrt{2/k})\tau
\label{eq:predlimit}
\eeq
as an approximate upper bound on the critical pattern extent $\Dphi_c$. Numerical data are in good agreement with our theoretical predictions $\Dphi_c^{\mathrm{glob.}}$ and $\Dphi_c^{\mathrm{max}}$ (cf.~\Figref{fig:2}).

To examine the stability of periodic firing patterns in the presence of heterogenous couplings $\eps_{ij}$, we consider a pattern
\beq
\phi_i(zT)=\phi_0+\Delta\phi_{i,0},\quad\quad z\in \mathbb{Z},
\label{eq:patterngeneral}
\eeq
of period $T$ where  $\phi_0$ represents a common reference phase and $\Delta\phi_{i,0}$ defines the relative phase shift of oscillator $i$ within the pattern. For simplicity, we order the firing events according to $\Delta\phi_{i,1}>\Delta\phi_{i,2}>\ldots>\Delta\phi_{i,k_i}$ where $\Dphi_{i,n}$ denotes the phase shift of the oscillator from which oscillator $i$ receives its $n$-th signal within a given period. For example, if $j'$ fires first of all oscillators connected to $i$, then $\Dphi_{i,1}=\Dphi_{j',0}$, cf.~\Eqref{eq:patterngeneral}.

By definition, all phases increase uniformly in time except for two kinds of discrete events: sending and receiving pulses. We follow the dynamics of a periodic state, $\phi_i(0)=\phi_i(T)$ for all $i$, event by event (cf.\ \cite{timme}) starting with a reference phase $\phi_0$ chosen such that all oscillators have fired but not yet received the generated pulses within a given period.  The collective period of the unperturbed firing pattern is given by
\beq
T=\tau+1+\Delta\phi_{i,0}-\Delta\phi_{i,k_i}-\sigma_{i,k_i}^{\Dphi}\ ,
\label{eq:natper}
\eeq
where $\sigma_{i,0}^{\chi}:=\tau$ and
\beq
\sigma_{i,n}^{\chi}:=\iU(U(\sigma^{\chi}_{i,n-1}+\chi_{i,n-1}-\chi_{i,n})+\eps_{i,n})
\label{eq:sigmadef}
\eeq
(here with $\chi=\Dphi$) are the recursively defined phases of oscillator $i$ right after having received the first $n$ of its $k_i$ inputs ($\eps_{i,1},\ldots,\eps_{i,k_i}$, ordered accordingly) in a given period.

Adding small perturbations $\delta_{i,0}(0)=:\delta_{i,0}$ satisfying
\beq
\max_i|\delta_{i,0}|<\frac 12 \Big(\min_{\substack{i,j\\ i\ne j}} |\Delta\phi_{i,0}-\Delta\phi_{j,0}|\Big )
\label{eq:deltaconstraint}
\eeq
to the phases of the oscillators $i$ before firing ensures preservation of the ordering of the firing events. Thus, denoting initial phases as $\phi_i(0)=\phi_0+\Dvphi_{i,0}$, where $\Dvphi_{i,0}:=\Delta\phi_{i,0}+\delta_{i,0}$, and sorting these relative phase shifts as before, $\Dvphi_{i,1}>\Dvphi_{i,2}>\ldots>\Dvphi_{i,k_i}$, it follows from $\Dphi_{i,k}<\Dphi_{i,l}$ that $\Dvphi_{i,k}<\Dvphi_{i,l}$ implying $\Dvphi_{i,n}=\Delta\phi_{i,n}+\delta_{i,n}$ for all $i$ and for all $n$. Following the evolution of the perturbed state event by event we obtain an expression for the time $T_i=\frac{\tau}2+1-\Dvphi_{i,k_i}-\sigma_{i,k_i}^{\Dvphi}$ to the next firing of $i$. This leads to a nonlinear period-$T$ map
\beq
\delta_{i,0}(T)=T-(T_i+\frac{\tau}2+\Delta\phi_{i,0})=\delta_{i,k_i}-\sigma_{i,k_i}^{\Dphi}+\sigma_{i,k_i}^{\Dvphi}
\label{eq:pertafterT}
\eeq
of the perturbations, where we used definition \Eref{eq:sigmadef} with $\chi=\Dphi$ and $\chi=\Dvphi$. Approximating
\beq
\sigma_{i,k_i}^{\Dvphi}\doteq\sigma_{i,k_i}^{\Dphi}+\sum_{n=1}^{k_i}(\delta_{i,n-1}-\delta_{i,n})p_{i,n-1}
\eeq
to first order in $\delta_{i,0}$, where $p_{i,n}:=\prod_{l=n}^{k_i-1}\Big(U'(\sigma_{i,l}^{\Dphi}+\Dphi_{i,l}-\Dphi_{i,l+1})/U'(\sigma_{i,l+1}^{\Dphi})\Big)$ for $n<k_i$ and $p_{i,k_i}=1$, yields
\beq
\delta_{i,0}(T)\doteq\sum_{j=1}^{N} M_{ij}\delta_{j,0}.
\label{eq:linop_simple}
\eeq
Here, the matrix $M_{ij}$ has diagonal elements $M_{ii}=p_{i,0}$, and the non-zero, off-diagonal elements are $M_{ij}=p_{i,n}-p_{i,n-1}$, where $j$ is the index of the oscillator sending the $n$-th signal to $i$ (i.e.\ $\Dphi_{j,0}=\Dphi_{i,n}$). Time translation invariance implies $\sum_j M_{ij}=1$ for all $i$. In addition, if either $U''>0$ and $\hat\eps>0$ or $U''<0$ and $\hat\eps<0$, then $p_{i,l}<p_{i,m}<1$ for $l<m$ which guarantees that all $M_{ij}\ge 0$. Using these properties of $\matrixer{M}$ it is straightforward to show that $\max_i |\delta_{i,0}(T)| \le \max_i |\delta_{i,0}(0)|$, which implies stability of the firing pattern.

Numerical simulations indicate that firing patterns are determined by the network topology and the structure of the heterogeneity. How does heterogeneity in a complex network control the precise timing of pulses that constitute a pattern? 

\begin{figure}
\centering
\dev{
\epsfig{file=figs/fig3ext.eps, width=0.9\columnwidth}}
\rel{\epsfig{file=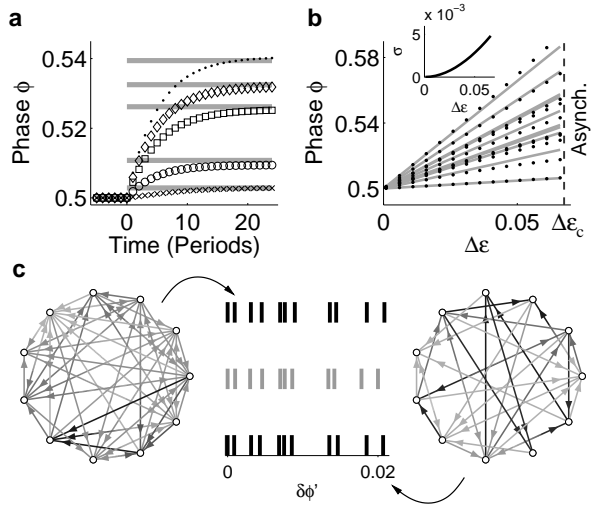, width=0.9\columnwidth}}
\caption{Predicting firing patterns and designing coupling architectures to create a prescribed pattern ($b=2, \hat\eps=-0.1, \tau=0.1$, predictions to first order). (a),(b) Comparison of prediction (grey lines) and simulation (markers) of selected phases of a pattern ($N=100$, $p=0.4$). (a) Phases shown period by period for $\Deps=0.25$. Heterogeneity is instantaneously introduced at time $t=0$. The attractor is reached asymptotically. (b) Pattern for increasing disorder $\Deps$ ($\matrixer{J}$ fixed). Inset: standard deviation of prediction from simulation. (c) Coupling strengths (indicated by shading) in two different heterogenous networks ($N=11$, left: $p=0.6$, right: $p=0.3$) designed to create a prescribed firing pattern (middle row, grey). Patterns in black are obtained by simulating these designed networks (initialized to the synchronous state).}
\label{fig:3}
\end{figure}

In the following we analytically predict the pattern resulting from the underlying coupling architecture. Assume that a given coupling architecture $\eps_{ij}$ induces a pattern \Eref{eq:patterngeneral} of period $T$. A change in the coupling matrix $\eps '_{ij}=\eps_{ij}+\deps _{ij}$ below the critical disorder leads to a new periodic state $\phi '_{i}(zT')=\phi _{i}(0)+\dphi_{i,0}$ of similar period $T'$ that is characterized by the phase shift $\vectorer{\dphi}=(\dphi_{1,0},\ldots,\dphi_{N,0})$. Defining $\DPhi_{i,n}:=\Dphi_{i,n}+\dphi_{i,n}$, we pick an arbitrary reference oscillator $l$ and adjust time such that its reference phase is $\DPhi_{l,0}=0$. The period of oscillator $l$ in the network defined by the couplings $\eps'_{ij}$ is then given by $T_l=1+\tau-\Dphi_{l,k_l}-\dphi_{l,k_l}-\sigma_{l,k_l}^{\DPhi}$ (where $\dphi_{i,j}$ is an ordering of $\dphi_{i,0}$ in analogy to the ordering $\Dphi_{i,j}$ of $\Dphi_{i,0}$). After the time $T_l$ the phase of an oscillator $i\ne l$ is given by
\beq
\phi'_i(T_l)=\sigma_{i,k_i}^{\DPhi}-\sigma_{l,k_l}^{\DPhi}+\Dphi_{i,k_i}-\Dphi_{l,k_l}+\dphi_{i,k_i}-\dphi_{l,k_l}.
\eeq
Periodicity of this state requires
\beq
R_{i}(\vectorer{\dphi},\deps):=\phi '_{i}(T_l)-\phi '_{i}(0)\stackrel{!}=0
\label{eq:selfconsist}
\eeq
for all $i$, which is an exact, implicit system of algebraic equations for $\vectorer{\dphi}$. The collective period $T'$ of the pattern is then given by $T_l$ evaluated at the actual phase shifts $\vectorer{\dphi}$. Expanding $R_i$ and $T_l$ to first order in $\deps_{ij}$ and $\dphi_{i,0}$ yields
\beq
\sum _{j} \frac{\partial R_{i}}{\partial \dphi_{j,0}}\Big \vert _{(0,0)} \dphi _{j,0}
=-\sum _{j}\frac{\partial R_{i}}{\partial \deps_{ij}}\Big \vert _{(0,0)}\deps_{ij}
\eeq
where $R_i(0,0)=\phi_i(T)-\phi_i(0)=0$ by definition. Thus $\vectorer{\dphi}$ can be approximated to first order by solving a system of $N$ linear equations. This first-order prediction yields a good approximation of the actual firing pattern for small disorder $\Deps$ (see, e.g.\ \Figref{fig:3}(a),(b)) and reasonably describes the qualitative relative firing times up to $\Deps=\Deps_c$ (cf.~\Figref{fig:3}(b)). 

The above method may be reversed to design heterogenous coupling architectures in order to create a prescribed pattern defined by $\vectorer{\dphi}$. Here, we fix the period $T'$ via the reference oscillator $l$, and solve each of the plane equations $i$ derived by linearizing \Eref{eq:selfconsist} with $T'=T_l$ for a suitable set of $\deps_{ij}$. Two examples of networks designed to exhibit the same prescribed pattern are shown in \Figref{fig:3}(c).

In summary, we have demonstrated that for small coupling disorder the synchronous state is replaced by precisely timed periodic firing patterns. We have shown how to predict the precise timing of pulses from a given coupling architecture, and reversely how to design networks in order to create prescribed patterns. There is a critical disorder strength, which we relate to the interaction delay $\tau$, beyond which only asynchronous, aperiodic states are observed.

Similar behavior is expected for other sources of heterogeneity, such as delay distributions, oscillator specific interactions $U_i(\phi)$ or heterogenous frequencies. The continuous transition found for small disorder is very different from that found previously for threshold-induced synchronization where oscillators are split into one synchronous and one asynchronous subpopulation due to heterogeneity \cite{tsodyks}. It has been shown recently that periodic firing patterns can be obtained and even learned in networks with global inhibition and no interaction delays as a perturbation of the asynchronous state \cite{spikepatterns}. In contradistinction, the patterns discussed in this Letter are not related to (but coexist with) asynchronous states, and emerge from the synchronous state due to heterogeneity. The simplicity of the original synchronous state aided us to clarify how heterogeneity of networks with a complicated structure controls their precise dynamics. These results are a first step towards understanding and designing the dynamics of real world networks that exhibit complex structure and heterogeneity.


\begin{thebibliography}{}
\bibitem{howmeasure} Numerical simulations start with phases initialized in the synchronous state \Eref{eq:synstate} of the homogenous network. The disorder $\Deps$ is increased (at fixed $J$) from $\Deps=0$ in increments of $10^{-3}$. For each value of $\Deps$, the phases are evolved for 300 firings of a reference oscillator, plotted, and used as initial conditions for the simulation at the next higher value of $\Deps$.
\bibitem{explainphasedist} Numerically we observe a unimodal phase distribution. The assumption of a uniform distribution, however, yields a good approximation (cf.~\Figref{fig:2}).
\end{thebibliography}
\end{document}